\documentclass[rnote]{aa}
\usepackage{fixltx2e,longtable,ifpdf,amsmath,amssymb,natbib}
\ifpdf
  \usepackage{aeguill}
  \usepackage[pdftex]{graphicx,color}
\else
  \usepackage[dvips]{graphicx,color}
\fi

\usepackage{etoolbox}
\AtBeginEnvironment{tabular}{\tiny}

\begin{document}

\title{Improved astrometry for the Bohannan \& Epps catalogue}

\author{Ian D. Howarth}
\institute{Dept. of Physics \& Astronomy, UCL, Gower St., London WC1E
  6BT, UK\\
\email{idh@star.ucl.ac.uk}}

\date{Submitted 17/8/2012; accepted 5/10/2012}

 
  \abstract
    {}
   {Accurate astrometry is required to reliably cross-match 20th-century catalogues
     against 21st-century surveys.   The present work aims to provide such
     astrometry for the 625 entries of the Bohannan \& Epps (BE74)
     catalogue of H$\alpha$ emission-line stars.}
   {BE74 targets have been individually
    identified in digital images and, in most cases, unambiguously matched to
     entries in the UCAC4 astrometric catalogue.}
   {Sub-arcsecond astrometry is now available for almost all
     BE74 stars.
   Several identification errors in the literature illustrate
   the perils of relying solely on positional coincidences using
   poorer-quality astrometry.}
   {}


   \titlerunning{Astrometry for BE74}
   \authorrunning{Ian D. Howarth}
   \maketitle
%

\section{Introduction}

Although the use of digital detectors and computer manipulation of
images is now ubiquitous, many pioneering surveys were conducted in
the days of photographic observations.  This is particularly
true of the Large Magellanic Cloud (LMC), where precedent ensures that most
of the brighter stars are still commonly identified by
catalogue numbers from surveys conducted in the photographic era, such as
the Henry Draper Extension (\citealt{Cannon36}; HDE), the
Radcliffe study by \citeauthor{Feast60} (\citeyear{Feast60}; R numbers\footnote{The CDS uses `RMC'
  to identify Radcliffe 
stars.}), and
work by \citeauthor{Henize56} (\citeyear{Henize56};
\mbox{LH$\alpha$120-S} identifiers for LMC emission-line stars)
and 
by \citeauthor{Sanduleak70} (\citeyear{Sanduleak70}; Sk
identifiers).

A difficulty confronting early authors was that the determination of
precise equatorial co-ordinates involved time-consuming manual
measurements with opto-mechanical plate-measuring machines, and
subsequent tedious calculations (as well as requiring a dense,
good-quality grid of reference stars).  This laborious effort was
invariably eschewed in favour of co-ordinates quoted to only
$\sim$arc{\-}minute precision and, in most cases, provision of supporting
finder charts.

In the modern era the identification problem is reversed: in
large-scale digital surveys, precise co-ordinates are quickly and
routinely obtained, but the task of visually checking many targets
against numerous published finding charts is discouragingly
burdensome.  Cross-identification based solely on co-ordinate
coincidences from crude astro{\-}metry
is not reliable in dense LMC starfields, but has nevertheless proven
enticing to a number of authors.  As a result, the literature is
littered with misidentifications based on approximate positional
matches, unverified by checks against original sources.

Heroic efforts by Brian Skiff at Lowell Observatory have greatly improved the
situation.  His work (unpublished, but incorporated into CDS
databases) includes $\sim$arc{\-}second astro{\-}metry for the HDE, Sk, and
LH$\alpha$ catalogues, based on careful examination of original sources.
There remain, however, two important, extensive surveys of bright H$\alpha$
emission-line stars in the LMC for which only arc{\-}minute
astro{\-}metry is available: the Lindsay surveys \citep{Lindsay63b,
Andrews64} and the Bohannan \& Epps study (\citealt{Bohannan74};
BE74).

Lindsay's papers give co-ordinates to the nearest arc{\-}minute, but
no finder charts.  Plausible identifications of many of
the $\sim$800 targets may be possible, based on positional and brightness
coincidences, but the absence of finder charts means
that \emph{secure} identifications are now not generally
feasible.\footnote{The authors provide cross-identifications with 126
  LH$\alpha$ objects; these sources are therefore reliably recoverable
  (through Henize's charts and Skiff's astro{\-}metry).}  Given the potential for errors, it
is the present author's opinion that the conservative position is to
consider Lindsay's stars lost to science, for the most part.

In contrast, \citet{Bohannan74} provided identification charts which
should allow secure identification of nearly all their emission-line
stars, and hence new astro{\-}metry with accuracy and precision suited
to cross-identifications in modern large-scale surveys.  The purpose
of this Note is to report such astrometry.

\begin{figure}
\centering
\includegraphics[scale=0.45,angle=-90]{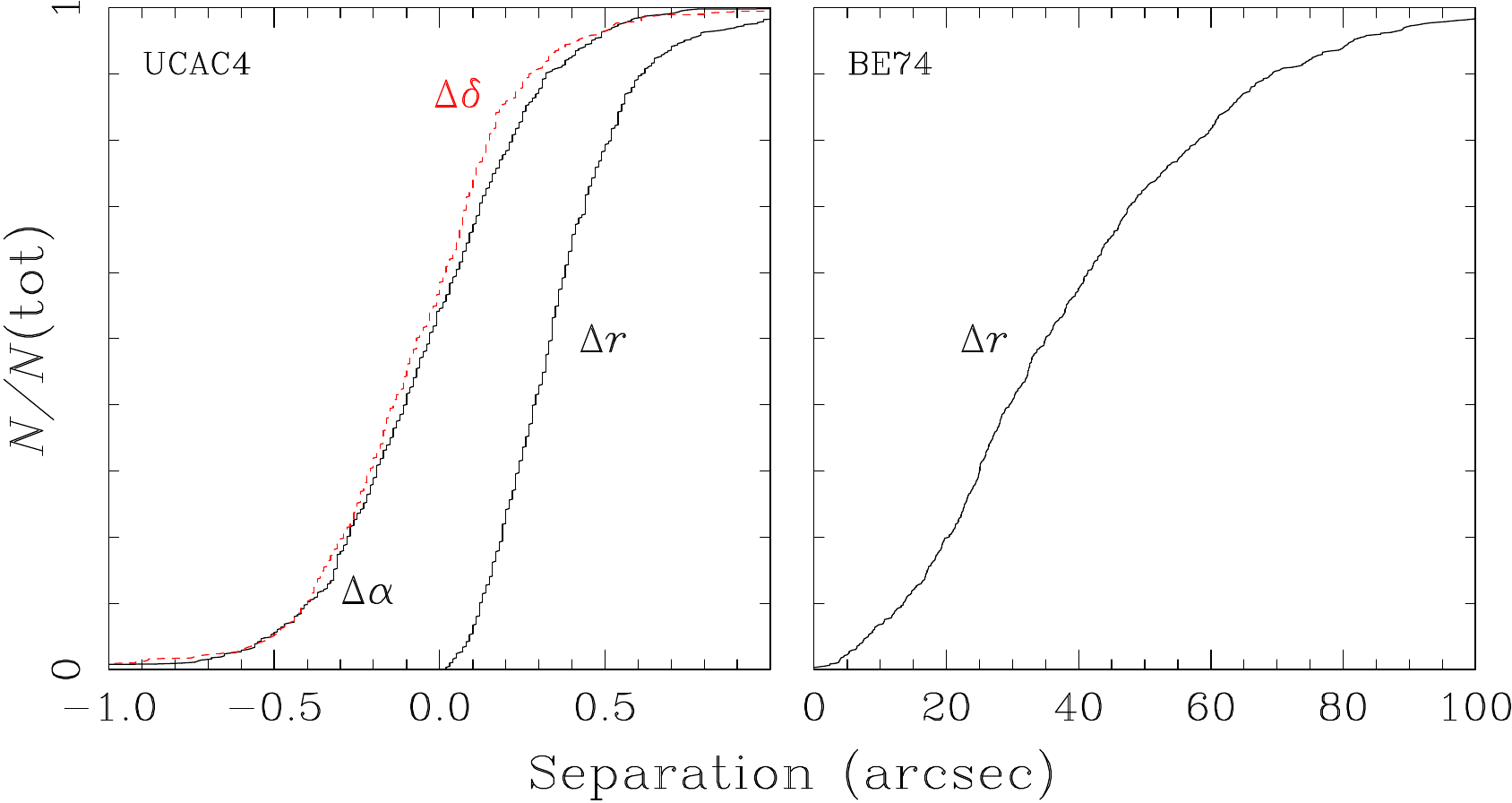}
\caption{Left panel: cumulative distribution functions for offsets
  between interactively measured co-ordinates and adopted UCAC4
  positions, for right ascension, declination, and absolute
  differences.  Right panel:  cumulative distribution function for
  offsets between BE74's positions and the current astrometry.}
\label{fig:one}
\end{figure}

\section{Methods}

\citeauthor{Bohannan74} identified their H$\alpha$ emission-line stars
on images from the \citeauthor{Hodge67} Atlas \citep{Hodge67}.  In
spite of the small plate scale, this allows subsequent secure
identification of nearly all stars in larger-scale digital images.  In
practice, this identification was normally carried out using CDS's
Aladin tool \citep{Bonnarel00} in conjunction with a much magnified
pdf copy of \citet{Bohannan74} from the NASA Astro{\-}physics Data
System. In a number of cases, the original \citeauthor{Hodge67} Atlas
was examined to resolve ambiguities.

Other than for a few bright targets, a semi-transparent window of the
DSS2 red image from Aladin was overlaid on the pdf, at matching
scales; where necessary, an image rotation was also applied, using the
GNU Image Manipulation Program.  In general, this allowed a positive
identification of the BE74 target with a single object on the DSS2
image, for which a position was recorded interactively, and
transferred to a data file by copy-and-paste.  (With distance moduli
of $\sim$18.5, proper motions are negligible for these sources, so
differences in epoch of observation are of no importance in this
context.)

Correlating the results against UCAC4 \citep{Zacharias12} gave a
positive match with a single target within 5$\arcsec$ in most cases,
with a small systematic offset: $\Delta\alpha = +0\fs11$, $\Delta\delta
= -0\farcs6$ (UCAC4$-$Aladin).  This offset is presumed to be due to
small errors in matching the DSS2 images to the ICRS reference frame,
so the interactively recorded measurements were corrected accordingly.
After applying this correction, a second pass was made against UCAC4
with a 2$\arcsec$-radius window to get final positions.  The results
are listed in the Table~1,\footnote{Tables 1--3 are available on-line
  at the CDS.} which is the main data product of this
Note.  Fig.~\ref{fig:one} (left-hand panel) shows that the positional
differences between corrected interactive measurements and UCAC4 are
Normally distributed, and are less than one arc-second in the great
majority of cases.

\section{Discussion}

The co-ordinates reported here are intended to establish precise
positions for the objects marked by BE74 on their finder charts,
independently of other investigations (largely to avoid any danger of
propagating existing misidentifications).  BE74 state that these
charts, rather than their published co-ordinates, best define their
targets, although a potential source of error is that they may not
always have matched the emission-line star from their objective-prism
plates with the correct object on the \citeauthor{Hodge67} Atlas (cf.,
e.g., BE74~602; Table~2).  Furthermore, in a number of cases multiple
bright sources are present on DSS2 images (and sometimes on the
\citeauthor{Hodge67} atlas) within the BE74 identification circle;
Table~2 discusses most of these instances.

Of course, in addition to ambiguities and possible errors in the
original materials, there is certainly also the potential for
misidentifications and mismeasurements in the present work.
Cross-checking the adopted positions against BE74's co-ordinates
initially disclosed six faulty results requiring correction in the
former (all believed to be due to failures to `copy' a correct
position before `pasting'), suggesting a residual error rate from this
source of better than 1\%.

Agreement between the finally adopted positions and BE74's original
co-ordinates is generally satisfactory, with matches to within
1--2$\arcmin$ (Fig.~\ref{fig:one}, right-hand panel), although some
discrepancies remain (cf.\ Table~2).  The current co-ordinates have
also been cross-matched to Skiff's astro{\-}metry for HDE, Sk, and
LH$\alpha$120-S sources, within a 5$\arcsec$ radius around the
interactively recorded positions (Table~3, on-line).  Where matches
are found, the positional differences are almost all sub-arcsecond
(essentially, the differences between UCAC2 and UCAC4 results),
giving confidence in the assigned correspondences.  In the few cases
where larger offsets occured, DSS2 imagery was re-examined;
invariably, the differences were found to arise because the BE74
object is not a single point source.

Finally, it is perhaps worth concluding with an explicit comment that,
while the positions reported here are precise, they may not
necessarily always be accurate, for the reasons just set out.  At the
least, for critical cases the results in Table~1 should be considered
in conjunction with the notes on individual sources given in Table~2.

\addtocounter{table}{1}
\begin{table*}
\caption{Notes on individual objects.  `BE74' refers to \citet{Bohannan74}; `HW'
  to charts from the \citeauthor{Hodge67} Atlas \citep{Hodge67};
  `RPs' to emission-line stars catalogued by \citeauthor{Reid12} (\citeyear{Reid12};
  RP12); 
  and `DSS2' to the red Digital Sky Survey images accessed through
  Aladin.  `A', `B' etc. refers to separate entries in Table~1.}             
\label{table:2}      
\centering          
\scalebox{0.91}{
\begin{tabular}{rp{19cm}}
\hline\hline       
BE74&Notes\\
\hline                    
  2  &NGC 1735 (cluster)\\

  37  &Identification uncertain.  
  The object marked by BE74 is the northerly of two
  stars visible on DSS2;  the companion is 6$\arcsec$ distant, SSE.  
  A brighter DSS2 star (not visible on the HW B chart, but clear on
  the V chart) is $\sim35\arcsec$ E.\\

 44  &Identification uncertain;  BE74 co-ords and finder-chart object
  differ by $3\farcm6$.  HDE~269504 (Sk~$-67$~100;  B0$\;$Ia
  according to \citealt{Fitzpatrick93}) is the brightest
  object at the published co-ords.\\

 51  &BE74 mark a triple object on the HW atlas\\

 54  &Assumed to be the brighter `A' (N) 
      component of a pair on DSS2 (unresolved in the HW atlas).\\

 62  &Elongated (multiple?) image on DSS2.\\

 68  &Assumed to be the middle (brightest) of three DSS2 stars.\\

 69  &Assumed to be the brightest (S) of three DSS2 objects.\\

 72  &SW component of multiple object (as per BE74 notes).\\

 74  &Wrongly identified with RPs1077 by RP12 (correct match is BE74~75/RPs1077).\\

 80  &MFH2006 Cl1 (cluster; \citealt{Martayan06})  \\

 84  &Brighter (E) component of double on DSS2.\\

109  &Recorded as 2 separate objects in UCAC4; barely resolved in DSS2.\\

127  &`A' is the brighter (W) component of a pair on DSS2 (unresolved in HW atlas).\\

129  &BSDL 2450  (cluster; \citealt{Bica99}).\\

139  &Identification uncertain;  BE74 finder chart and co-ords differ by $8\farcm5$.  BE74's quoted
co-ords are in RA sequence (as is usual), but the marked object is out of
RA order, and is therefore suspect.   The nearest moderately bright star to the published co-ords is 
UCAC4~116-009443.\\

143  &Assumed to be brighter W component of pair on DSS2 (unresolved
in HW atlas; companion 2$\arcsec$ E).\\

159  &BE74 identify 159 with the Wolf-Rayet
star WS4=BAT99-8 \citep{Westerlund64, BAT99},
but this is not the star marked on the BE74 finder chart
\citep{Fehrenbach76}, which does, however, match LH$\alpha$120-S70.\\

162 &Wrongly identified with RPs1757 by RP12 (RPs1757 is much fainter).\\

168  &Elongated image on DSS2, probably multiple.\\

171  &N component of double on DSS2 (secondary is 6$\arcsec$ SW,
unresolved on the BE74 chart).
      Identification uncertain; BE74 identify 171 with 
      WS6=BAT99-11 \citep{Westerlund64, BAT99},
      but this is not the star marked on the BE74 finder chart \citep{Fehrenbach76}.\\

191  &Elongated (multiple?) image in DSS2.\\

197  &Double on DSS2 (unresolved on the BE74 chart).\\

207  &Centre object of three (as in BE74 notes).\\

223  &BE74 co-ords wrongly duplicate the entry for BE74~224 (7$\arcmin$ to the south).\\

227  &Typographical error in RP12;  correct match is BE74~277/RPs2160, not BE74~227/RPs2160.\\

232  &Brighter SW component of double on DSS2.\\

238  &BE74 finder-chart object is 4$\arcmin$ N of BE74 co-ords.\\

242  &Wrongly identified with RPs1374 by RP12 (correct match is BE74~242/RPs1373).\\

246  &Unresolved double in the HW atlas.\\

276  &Southern object of pair is brighter on DSS2, and matches RPs988 
     (BE74:  ``Impossible to tell from which emission arises'').  RP12 also
     (wrongly) identify BE74~276 with RPs989.\\

277  &Elongated image on DSS2?\\

292  &ID uncertain on BE74 atlas.\\

294  &S component of double on DSS2.\\

299  &NGC~1994 (HDE~269599, cluster).\\

360  &Object marked on BE74 finder chart is 6$\arcmin$ N of BE74 co-ords.\\

380  &ID uncertain on BE74 atlas.\\

383  &HDE~269828 (cluster).\\

394  &Not marked on BE74 charts; several possible candidates.\\

400  &Close double, emission-line (WR) star is W component
(Sk~$-69$~223, BAT99-85; \citealt{Prevot81}).\\

402  &KMK88 91 (cluster; \citealt{Kontizas88}).\\

426  &Wrongly identified with RPs286 by RP12 (cf.\ their Fig.~4; RPs286 is much fainter).\\

434  &Not marked on BE74 charts; several possible candidates.\\

439  &Not marked on BE74 charts; two plausible possibilities measured (439a is HV 2774).\\

441  &Elongated (multiple?) image on DSS2.\\

443  &BE74 finder chart ambiguous.   Wrongly identified with both
      RPs236 and RPs237 by RP12.\\

445  &KMHK 1230 (multiple; \citealt{Kontizas90}).\\

451  &Wrongly identified with RPs1023 by RP12 (RPs1023 is much fainter).\\

477  &BE74 finder-chart object matches HDE~268687/Sk~$-$69~13
(F6$\;$Ia according to \citealt{Ardeberg72}), but is 4$\arcmin$ N of BE74 co-ords.\\

487  & Two stars in BE74 identifying circle.\\

494  &`A' is the object shown on BE74 chart;  `B' is not visible in HW but is
brighter on DSS2 (and both are within the BE74 ID circle).\\

497  &Two bright stars (septn. $6\farcs8$) unresolved on BE74 chart;  `A' (HD~32763) is the brighter object in DSS2, and 
matches Skiff astrometry for LH$\alpha$120-S149/Sk~$-$70~29.\\

499  &Brightest (NW) object of three within the circle marked by BE74.\\

509  &Distorted DSS2 image; two close objects in UCAC3 match this distortion.\\

518  &BE74:  ``Perhaps emission arises from 
      S and brighter member of pair'' (agrees with Skiff
      LH$\alpha$120~S156 astro{\-}metry).\\

520  &Uncertain ID (no object clearly visible on the HW B chart).\\

522  &BE74 finder object is $7\farcm5$ S of BE74 co-ords.\\

528  &Elongated (multiple?) image on DSS2.\\

565  &Identified with RPs1344 by RP12, but this is a different
      object to that marked by BE74 (16$\arcsec$ distant).  However, RPs1344 is
      only $\sim$0.5m fainter than BE74~565;  possible misidentification on BE74 chart?\\

567  &NW (brighter) `A' component has elongated (multiple?) image on
DSS2.  `B' component matches Skiff LH$\alpha$120-S102 astro{\-}metry.\\

570  &Misidentified as BE571 in Simbad at the time of writing.\\

574  &Identified with RPs870 by RP12, but this is a different object
     to that marked by BE74 (12$\arcsec$ distant).    However, the object
     marked by BE74 is very faint on DSS2, while RPs870 is almost
     invisible on HW;  possible misidentification on BE74 chart?\\

578  &SE, `A' component is the only one of three DSS2 objects visible
      in the BE74 chart.  `B' component matches Skiff LH$\alpha$120-S104 astro{\-}metry.\\

579 &Wrongly identified with both RPs873 and RPs874 by RP12 (both
     RPs objects are much fainter).\\

581 &Identified with RPs886 by RP12;  this is $1\farcm3$ S of the marked BE74 position, but 
     the BE74 object is very faint on DSS2.  Possible misidentification on BE74 chart?\\

582 &  Two objects are marked as `587' on BE74 chart; N object is
     actually BE74~582.\\

595 &Wrongly identified with RPs443 by RP12 (correct match is BE74~596/RPs443).\\

602  &Nova Mensae 1970b.  Co-ordinates reported in Table~1 refer to
the finder-chart object
     marked by BE74, but this is \emph{not} the nova.\newline  602X in Table~1
     is the approximate position of the nova.\\

620  &MHW2005 1145 \citep{Meynadier05}; emission comes from the nebular `blob'.\\

\hline                  
\end{tabular}
}
\end{table*}

\acknowledgements{I am grateful to Brian Skiff for correspondence;
  Jean Guibert for useful suggestions; and
  the BBC for its broadcast coverage of the 2012 Olympics, which
  facilitated the measurements reported here.  The work relied on tools
  provided by the GNU/Linux open-source community, NASA's
  Astro{\-}physics Data System and, particularly, the CDS, Strasbourg;
  without these resources, this study would not have been possible.}

\bibliography{IDH}

\bibliographystyle{aa}

\end{document}